# A Reputation System for Marketplaces - Viability Assessment

**Anton Kolonin, Ben Goertzel, Cassio Pennachin, Deborah Duong, Marco Argentieri, Nejc Znidar**

SingularityNET Foundation, Amsterdam, Netherlands

{anton, ben, cassio}@singularitynet.io

## Abstract

In this work we explore the implementation of the reputation system for a generic marketplace, describe details of the algorithm and parameters driving its operation, justify an approach to simulation modeling, and explore how various kinds of reputation systems with different parameters impact the economic security of the marketplace. Our emphasis here is on the protection of consumers by means of an ability to distinguish between cheating participants and honest participants, as well as the ability to minimize honest participant's losses due to scam.

## 1 Introduction

The importance of automated reputation assessment in online communities has been the focus of several studies over the past decade, as it can be found in the earlier works of the other authors: [Zheng and Jin, 2009], [Swamynathan et al., 2010], [Sänger and Pernul, 2018]. This topic is particularly interesting given the growing interest in distributed interaction systems and marketplace platforms, where suppliers and consumers are connected on a peer-to-peer basis [Gupta *et al*., 2003]. The ability to build reliable reputation systems gets especially critical in the case of densely populated marketplaces with anonymity of participants [Androulaki *et al*., 2008]. In such marketplaces the trustability of most of one's peers involved in interactions can not be verified other than on basis of personal experience, so that public reputation systems based on collaborative feedback can potentially help [Blömer *et al*., 2015].

The scope of our work extends ideas relying on graph analysis where oriented social graphs [Garin and Mescheriakov, 2019] make it possible to extract social relationships between the participants of the society or marketplace and assess trustworthiness of each of the participants based on the volume of social experiences in the market environment [Kolonin, 2019].

Given the generic design of a reputation system for artificial and online societies [Kolonin *et al*., 2018] presents an overall framework for building reputation systems, in this work we focus on implementing a simplified version serving the marketplace environment and on adjusting system parameters suitable for this case. For the latter purpose, we design and perform simulation modeling that would enable one to figure out the appropriate set of parameters for maximizing the system's ability to satisfy its requirements. The key principle of the system remains the "weighted liquid rank" algorithm [Kolonin *et al*., 2018] which defines a computation of reputation based on explicit ratings being exchanged with by market participants, weighted by values of respective financial transactions and taking into account the reputations of the raters issuing the ratings themselves.

The primary requirement of the reputation system is assumed to be to provide the ability to identify scamming agents that cheat the reputation system by pumping their reputations with worthless fake services, that is, it should provide "security". Secondarily, for the agents that are asserted as fair participants, the reputation system is required to provide the ability to assess their level of service appropriately, having it correctly represented in a reputation rank, that is, it should grant "equity". As part of granting "equity", the system should make it possible for novel fair agents entering the system to be recognized and not be suppressed by system, so as to grant "openness".

## 2 Reputation System Implementation and Parameters

In the scope of this work we substantially simplify the original design [Kolonin *et al*., 2018] of the reputation system, leaving multiple domains and dimensions out of the scope of consideration and abstracting from different implementation options. Here we focus on implementation details of the "weghted liquid rank" algorithm and the multiple algorithm configuration parameters to be explored during the simulation modeling.

### 2.1 Reputation System Overview

In this work, we have explored reputation systems that may consume either implicit ratings $Q_{ij}$ corresponding to financial transactions made by agent *i* to agent *j* as or explicit ratings $F_{ij}$ between agents that may be optionally weighed with financial values of corresponding transactions conducted between the agents.

The rating system for explicit ratings was designed with a continuous rating scale defined in range between *0.0* and *1.0*, which could be mapped to range *0%* to *100%* with *0.0* or *0%* meaning complete dissatisfaction and *1.0* or *100%* standing for complete satisfactions. For practical applications and simulations of the reputation system it could be used with discrete rating values of "five-star" rating system, where *0,0, 0.25, 0.5, 0.75* and *1.0* would imply unsatisfactory, satisfactory, good, very good, end excellent, respectively. Alternatively, it could be thought in terms of possible downrating or negative rating indicated with value *0.0* while the remaining four values of *0.25, 0.5, 0.75* and *1.0* would represent "four-star" system of positive rating.

The implementation of the reputation system used in this work is based on the incremental computation model, so that a reputation state computed at the end of a previous observation period is used to compute a differential reputation update during subsequent observation periods. The previous reputation state is then blended with the differential update into the new reputation state. Each reputation state contains the reputation ranks of all agents known by the end of a given observation period.

### 2.2 Reputation System Parameters

In the implementation of the reputation system that we have explored, the following parameters were used:
- $R_d$ - default initial reputation rank in range *0.0* to *1.0* - given to newcomer agents initially;
- $R_c$ - decayed reputation in range *0.0* to *1.0* - to be reached by agents eventually if they have no ratings;
- $F_d$ - default explicit rating to assume if it no rating is left by rater, in range *0.0* to *1.0*;
- *C* - conservatism in range *0.0* to *1.0* as a blending ("alpha") factor for mixing previous reputation rank recorded at the beginning of the observed period and a differential one obtained during the observation period;
- *P* - precision as a decimal number like *10.5* or *0.01*, used to round/up or round down financial values or weights as *value = Round(value/precision)*;
- *Weighting* - with this boolean option set to *True,* the reputation system weighs explicit ratings with normalized financial values ( *0.0* to *1.0*) , and with it set to *False* it leaves the explicit ratings unweighted;
- *FullNorm* - with this boolean option set to *True* the system performs a full-scale normalization of incremental ratings, and with it set to False it has limited normalization, without lowering the minimum values to zero level;
- *Liquid* - with this boolean option set to *True* the reputation system accounts for rank of the rater when computing incremental reputation updates;
- *LogRanks* - with this boolean option set to *True* the reputation system applies the *log10(rank)* function to reputation ranks during normalization;
- *LogRatings* - with this boolean option set to *True* the reputation system applies *log10(1+value)* to financial values used for implicit ratings and financial weights used for weighted explicit ratings;

- *Aggregation* - with this boolean option set to *True* the reputation system aggregates all explicit ratings between each unique combination of two agents by computing the weighted average of ratings across the observation period;
- *Downrating* - with this boolean option set to *True* the reputation system translates original explicit rating values in to range *0.0-0.25* to negative values in the range *-1.0* to *0.0* and original values in the range *0.25-1.0* to the interval *0.0-1.0*.
- *UpdatePeriod* – the number of days to update the reputation state, considered as the observation period for computing incremental reputations.

## 2.3 Reputation System Algorithm and Implementation

The following describes the implementation of the "weighted liquid rank" algorithm [Kolonin *et al.*, 2018] used in the present implementation of simulation system described later in this paper.

1. For each observation period end date *t*, the reputation state at the last day *t-1* of the previous observation period is retrieved as a list of reputation ranks $R_{it-1}$ of every agent *i* at day *t-1*.

2. Either implicit financial ratings or explicit ratings for the observation period between *t-1* (exclusively) and *t* (inclusively) is retrieved.

3. Optionally, if the *Aggregation* option is set to *True*, all $N_{ij}$ ratings per observation period per each unique combination of rater agent *i* and rated agent *j* are averaged, so that the average implicit financial rating $Q_{ij}$, and unweighted explicit rating $F_{ij}$ may be computed for further use, in order to prevent excessive uprating or downrating between specific agents.

   $Q_{ij} = \Sigma_t Q_{ijt}/N_{ij}$ , in case of implicit financial ratings

   $F_{ij} = \Sigma_t F_{ijt}/N_{ij}$ , in case of unweighted explicit ratings

4. Optionally, if the *Precision* parameter is set to a value other than *1.0*, the financial values of the implicit or explicit ratings are re-scaled with $Q_{ij} = Round(Q_{ij} / Precision)$.

5. Optionally, if the *LogRatings* option is set to *True*, financial values of the implicit or explicit ratings are scaled to logarithmic scale as $Q_{ij} = If(Q_{ij} < 0, - log10(1 - Q_{ij}), log10(1 + Q_{ij}))$, where negative value may be corresponding to the case of transaction withdrawal or cancellation.

6. Optionally, if the *Downrating* option is set to *True*, the explicit ratings are translated from range *0.0 - 1.0* to range between *-1.0 and +1.0*, based on threshold value 0.25, so that $F_{ij} = If( F_d < 0.25, (F_d - 0.25)/0.25, (F_d - 0.25)/0.75 )$.

7. For the volume of the ratings for the observation period, either aggregated or not aggregated, the differential reputation update $dR_i$ for every agent *i* is computed in one of the three ways, for implicit financial ratings, unweighted explicit ratings or weighted explicit ratings, respectively, as follows. In each of the following cases, if the *Liquid* parameter is set to *False*, the reputation of the rater for the previous observation period is disregarded so that $R_{jt-1} = 1$. If the *Liquid* parameter is set to *True* but the reputation of the rater for the previous period is unknown yet, it as assumed to be default rank as $R_{jt-1} = R_d$. If the explicit rating is not known for given transaction between agents, it is expected to be default rating $F_{ij} = F_d$:

$dR_{it} = \Sigma_j (Q_{ij} * R_{jt-1})$, in case of implicit financial ratings;

$dR_{it} = \Sigma_j (F_{ij} * R_{jt-1})$, in case of unweighted explicit ratings;

$dR_{it} = \Sigma_j (F_{ij} * Q_{ij} * R_{jt-1})$, in case of weighted explicit ratings.

8. The differential update $dR_{it}$ for the observation period *t* is further normalized to range between *0.0* and *1.0* with optional logarithmic scaling to $ldR_{it} = log10(1 + dR_{it})$, if *LogRanks* is set to *True,* and optional full-scale normalization, if *FullNorm* is set to *True*:

   $ndR_{it} = dR_{it} / Max(dR_{it})$, if *LogRanks=False* and *FullNorm=False*;

   $ndR_{it} = (dR_{it} - Min(dR_{it})) / (Max(dR_{it}) - Min(dR_{it}))$, if *LogRanks=False* and *FullNorm=True*;

   $ndR_{it} = ldR_{it} / Max(ldR_{it})$, if *LogRanks=True* and *FullNorm=False*;

   $ndR_{it} = (ldR_{it} - Min(ldR_{it})) / (Max(ldR_{it}) - Min(ldR_{it}))$, if *LogRanks=True* and *FullNorm=True*;

9. The ranks from the earlier observation period are blended with normalized differential ranks with taking into account conservatism, as shown below. In the case that the reputation rank for the previous observation period $R_{it-1}$ is not known, default $R_d$ is used instead. In the case that $R_{it-1}$ is known but there is no update $ndR_{it}$ observed in given period, $R_c$ is used instead. That means, the reputation curve of the agents would start with $R_d$ and approach $R_c$ on inactivity eventually as an average across agents, yet further normalization adjusts reputations across all agents to the uniform range so the actual numbers may be different.

   $R_{it} = R_{it-1} * C + ndR_{it} * (1 - C)$

10. The computed blended reputation ranks are finally normalized to the range between *0.0* to *1.0* and stored to be used for computing reputations at the end of the next observation period.

In the scope of the project, the reference implementation of the reputation system has been initially done within the scope of open source Aigents Java project as the Reputationer class. For overall implementation details see *Reputationer* class in https://github.com/aigents/aigents-java/ and for blending and normalization functions refer to Summator class in the same project. The Aigents Java implementation of reputation system is provided with a Python adapter for SingularityNET reputation system prototype (see the *aigents_reputation_api.py* script in the https://github.com/singnet/reputation/ project). The latter project is also hosting further porting of the system to native Python implementation (see the *reputation_service_api.py* script).

## 3 Reputation System Simulation

In order to explore the reputation system design described above we have performed an initial round of simulation modeling. The simplistic simulation was used to perform a cursory exploration of the market conditions and parameters of the reputation system using simplified market behavior, to see if the reputation system can work on certain market conditions or not, what kind of the reputation system provides the best "security" and "equity", and what the optimal system parameters are.

## 3.1 Simulation Requirements

In our simulations, it was expected that amount of "good" agents with "honest" behaviors in the market is *80%* of the population and the amount of "bad" agents committing "scams" are the remaining *20%* so that the market was, in a sense, "overpopulated" by scammers. During the simulations, it was also assumed that the amount of transactions committed by "bad" agents is somewhat higher than the amount of transactions of by the "good" agents, so the "bad" agents are cheating by having the reputations of "fake" suppliers pumped up by "fake" consumers. It was also assumed that an average "bad" agent spends substantially less on every transaction, compared to an average "good" agent who, being honest, are actually making real purchases while scamming agents are spending money just in order to fool the reputation system. In the described study, in all simulations it was assumed that any of "bad" agents emits *10* times more transactions a day than the good one, but the ratio of average transaction payment amount varied. For the latter ratio, we considered three typical conditions: an "unhealthy" market with a ratio equal to *10* (so that the "good" agents have just *10* times more costly purchases), a "semi-healthy" one with a ratio of *20* and a "healthy" one with a ratio of *100*.

We have explored different ratios between amounts of "good" agents performing "honest" behaviors in the market and "bad" agents committing "scams". The following different ratios between percentages of consumers and suppliers were considered: 1) everyone is consumer and supplier, 2) *50%* are consumers and *50%* are suppliers, 3) *80%* are consumers and *20%* are suppliers, 4) *90%* are consumers and *10%* are suppliers. The whole range of ratios was explored in the course of initial simulation studies and the latter one (*90%* consumers per *10%* suppliers) was used for fine-tuning of system parameters in a way that corresponds to real-world marketplaces, as we have discovered in our studies in the field. We have studied real world marketplaces and what is a share of buyers vs sellers there. In our study we took into account 16 mature marketplaces for which we were able to collect data on the ratios of buyers vs sellers. This data was collected from publicly available information that those companies posted or the analysts estimated. These are all online marketplaces, ranging from huge marketplaces in broad markets such as Amazon, to very niche marketplaces such as Algorithmia. We then assessed those ratios based on how reliable they are (number of buyers and sellers might have had inaccuracies, such as counting inactive buyers, etc.) and how similar the niche is to ours (an AI as a service marketplace). We found this ratio *9:1* (9 times as many buyers than sellers) to be the most realistic.

The behavior of the "good" agents was such that any "good" consumer performs a random or semi-random search for suppliers and provides semi-random "positive" rating in the range of *0.25* to *1.0* to a "good" supplier. However, whenever a "bad" agent is selected by a "good" agent, the former is rated with "negative" *0.0* rating and never selected for purchase again.

The behavior of the "bad" agents was such that "bad" consumers provide fake "positive" *1.0* ratings to "bad" suppliers and never give a rating to "good" suppliers.

All simulations were performed in four different modes: 1) consumers do not use the reputation system to make a choice for supplier, so they pick either randomly or based on their own previous experiences; 2) consumers use a "traditional" reputation system running on the basis of explicit ratings that are not weighted by financial transaction costs; 3) consumers are using a reputation system running on the basis of explicit ratings weighted by financial transaction costs; 4) consumers are using a reputation system running on the basis of implicit financial ratings, which are financial transaction costs.

Different scales of the simulations were explored, for the market parameters and behavioral patterns above - 10 agents interacting during 10 days, 100 agents interacting during 3 months and 1000 agents interacting during 6 months.

The simplistic simulation, which code is available under the https://github.com/singnet/reputation/, project in the *reputation_scenario.py* script was assuming a completely random selection of suppliers by consumers, with only two exceptions. First, the "bad" suppliers experienced by "good" consumers in person were never selected by the latter again so the bad experiences were remembered by each of the honest agents. Second, when "good" consumers used the reputation system to select suppliers for purchase, with the choice restricted to only suppliers with computed reputations exceeding a reputation rank value equal to or above the 40% mark.

### 3.2 Simulation Performance Metrics

We have examined a few kinds of performance metrics for the reputation system. The first kind of metric measures the effect of the reputation system on the economy. The financial metrics serve to identify the amount of financial resources that fair consumers lose to scams, or the relative amount that scammers would earn by cheating the reputation system so they can be referenced to evaluate the economic value of a reputation system. The second kind of metric are just technical measures of the performance of the reputation system in terms of its accuracy. The latter metrics assess the degree to which we can identify scammers, assign appropriate ratings to good agents or solving both problems at the same time.

We present two metrics of financial kind: $C_{gbg}$ - "loss to scam" - a ratio of volume of transactions spent by good agents $V_{gb}$ to bad agents over the entire volume of the spending by all good agents $V_g$. Second, we have $C_{gbb}$ - "profit from scam" a ratio of the volume of bad agent earning transactions from good agents $V_{gb}$ over the spendings of bad agents $V_b$. The working reputation system would be expected to minimize both $C_{gbg}$ and $C_{gbb}$, the "fair" market's loss to scam and the profitability of scam.

$C_{gbg} = V_{gb} / V_g$ , "loss to scam"

$C_{gbb} = V_{gb} / V_b$ , "profit from scam"

We also explore metrics that assess the accuracy of the reputation system's prediction of whether each agent is good or bad, and ability of the reputation system to have a high accuracy for both assessments. These metrics compare distributions of expected reputations $R_{ea}$ per agent $a$ against lists of computed reputations $R_{ca}$ for the same set of agents. The scale of $R_{ea}$ is used to assess the reputation system's ability to predict "bad" agents, such that bad agents are expected to have $R_{ea} = 0$, and good agents are expected to have $R_{ea} = 1$.

The Pearson correlation coefficient $PCC(R_{ea}, R_{ca})$ between $R_{ea}$ and $R_{ca}$, different forms of accuracy for good, bad and mean as well as root-mean-square deviation metrics have been used to compare $R_{ca}$ against $R_{ea}$, as follows. Using these metrics, reputation system would be expected to provide values above zero at least, with values above *0.5* as a reasonably good result and values close to *1.0* as excellent results.

The accuracies for evaluating "good" agents $A_g$ and "bad" agents $A_b$ were defined as follows, assuming that the $R_{ca}$ and $R_{ea}$, are both defined on interval from *0.0* to *1.0*, representing "goodness" with good agents ultimately having a value of *1.0* and bad agents ultimately having a value of *0.0*. Respectively, we can consider $\tilde{R}_{ca} = 1 - R_{ca}$ and $\tilde{R}_{ea} = 1 - R_{ea}$, representing "badness" with the inverse meaning. In the following, we essentially compute weighted average values of computed "goodness" and "badness", with expected values used as averaging weights.

$A_g = \Sigma_a (R_{ca} * R_{ea}) / \Sigma_a (R_{ea})$

$A_b = \Sigma_a (\tilde{R}_{ca} * \tilde{R}_{ea}) / \Sigma_a (\tilde{R}_{ea})$

The metrics above would let us maximize the value of the system's ability to correctly identify the "good" and the

"bad" independently, and the following value of mixed average $A_m$ could let us measure the extent of maximizing the both.

$$A_m = (A_g + A_b) / 2$$

As a possible alternative for accuracy metrics above, we have used root-mean-square deviations, defined separately for "goodness" $D_g$ and "badness" $D_b$ and mixed $D_m$ as in the case of the accuracies above. These metrics give the level of error in identifying different sorts of agents correctly, with *1.0* being the worse and *0.0* being the best. The mixed deviation, $D_m$, represents a traditional error measure while specific $D_g$ and $D_b$ are computed with weighted averages of square deviations with expected "goodness" and "badness" values as weights, respectively.

$$D_m = \sqrt{(\Sigma_a((R_{ca} - R_{ea})^2) / N )}$$

$$D_g = \sqrt{(\Sigma_a((R_{ca} - R_{ea})^2 * R_{ea}) / \Sigma_a(R_{ea}) )}$$

$$D_b = \sqrt{(\Sigma_a((\tilde{R}_{ca} - \tilde{R}_{ea})^2 * \tilde{R}_{ea}) / \Sigma_a(\tilde{R}_{ea}) )}$$

## 3.3 Simulation Results

The following results were obtained and analyzed in order to explore the market conditions which make the reputation system work at all, as well as the combinations of parameters that make it work to greater extent.

| | good/bad volume ratio | Not using reputation system | | | | | | | | | | If using reputation system | | | | | | | | | |
|---|---|---|---|---|---|---|---|---|---|---|---|---|---|---|---|---|---|---|---|---|---|
| | | profit from scam, % | loss to scam, % | Pearson average | Pearson latest | avg good acc. | avg bad acc. | avg bal. acc. | avg good rmsd | avg bad rmsd | avg rmsd | profit from scam, % | loss to scam, % | Pearson average | Pearson latest | avg good acc. | avg bad acc. | avg bal. acc. | avg good rmsd | avg bad rmsd | avg rmsd |
| r_10_0.1 N/W | 4 | 4% | 1.1% | -1.00 | -1.00 | 0.09 | 0.00 | 0.05 | 0.91 | 1.00 | 0.93 | 4% | 1.1% | -1.00 | -1.00 | 0.09 | 0.00 | 0.05 | 0.91 | 1.00 | 0.93 |
| r_20_0.1 N/W | 8 | 9% | 1.1% | -1.00 | -1.00 | 0.09 | 0.00 | 0.05 | 0.91 | 1.00 | 0.93 | 9% | 1.1% | -1.00 | -1.00 | 0.09 | 0.00 | 0.05 | 0.91 | 1.00 | 0.93 |
| r_100_0.1 N/W | 40 | 44% | 1.1% | -1.00 | -1.00 | 0.09 | 0.00 | 0.05 | 0.91 | 1.00 | 0.93 | 44% | 1.1% | -1.00 | -1.00 | 0.09 | 0.00 | 0.05 | 0.91 | 1.00 | 0.93 |
| r_10_0.1 | 4 | 4% | 1.1% | -1.00 | -0.95 | 0.40 | 0.00 | 0.20 | 0.60 | 1.00 | 0.70 | 3% | 0.8% | 0.22 | 0.24 | 0.42 | 0.82 | 0.62 | 0.75 | 0.18 | 0.67 |
| r_20_0.1 | 8 | 9% | 1.1% | 0.99 | 0.81 | 0.94 | 0.62 | 0.78 | 0.06 | 0.38 | 0.18 | 5% | 0.7% | 0.63 | 0.59 | 0.78 | 0.84 | 0.81 | 0.41 | 0.16 | 0.37 |
| r_100_0.1 | 40 | 44% | 1.1% | 1.00 | 0.99 | 0.99 | 0.97 | 0.98 | 0.01 | 0.03 | 0.02 | 4% | 0.1% | 1.00 | 1.00 | 0.98 | 0.97 | 0.98 | 0.02 | 0.03 | 0.02 |
| p_10_0.1 | 4 | 4% | 1.1% | -0.85 | -0.04 | 0.88 | 0.01 | 0.45 | 0.12 | 0.99 | 0.46 | 4% | 1.0% | 0.19 | 0.21 | 0.26 | 0.93 | 0.59 | 0.86 | 0.07 | 0.77 |
| p_20_0.1 | 8 | 9% | 1.1% | 1.00 | 0.98 | 0.97 | 0.85 | 0.91 | 0.03 | 0.15 | 0.07 | 7% | 0.9% | 0.48 | 0.50 | 0.66 | 0.90 | 0.78 | 0.57 | 0.10 | 0.52 |
| p_100_0.1 | 40 | 44% | 1.1% | 1.00 | 1.00 | 0.99 | 0.93 | 0.96 | 0.01 | 0.07 | 0.03 | 36% | 0.9% | 0.82 | 0.84 | 0.92 | 0.92 | 0.92 | 0.27 | 0.08 | 0.25 |

Figure 1. Using or not using reputation systems based on different kinds of ratings for fixed system reputation system parameters with different market conditions. Columns 3-12 –reputation ranks for metrics are computed aside and not communicated to agents, columns 13-22 – agents use a reputation system to choose suppliers by consumers. Rows 1-3 – the reputation system is based on explicit ratings, not weighted by financial values, rows 4-6 – the reputation system is based on explicit ratings, weighted by financial values, rows 7-9 – the reputation system is based on implicit ratings, provided by financial values. Rows 1,4,7 – transactions made by good agents are 10 time more valuable than ones spent by bad agents, rows 2,5,6 – transactions by good agents are 20 times more valuable, rows 3,6,9 – transactions by good agents are 100 times more valuable. Reputation system parameters are: default = 0.5, conservativity = 0.5, decayed = 0.0, full normalization = true, logarithmic ratings = false, downrating = false. The coloring is used to outline ranges. Green means a reasonable range, i.e. strong positive connections between expected and obtained reputations (above 0.75 for PCC and accuracy and below 0.25 for RMSD). Yellow means a semi-reasonable range, i.e. positive connections between expected and obtained reputations (between 0.25 and 0.75 for Pearson and between 0.5 and 0.75 for Accuracy, between 0.5 and 0.25 for RMSD and scam profitability). Orange means an unreasonable range, i.e. no connection between expected and obtained reputations (below 0.25 for Pearson and below 0.5 for Accuracy, above 0.5 or RMSD and scam profitability). Note that the reasonable (green) range for loss to scam is below 1% and for profit from scam it is below 10%.

For a reputation system, to be working, we expect it to be able to solve two goals at once. Primarily, it would make it possible to identify "bad" cheating suppliers supported by cheating consumers so the that the overall loss that "good" agents making purchases suffer would be brought to minimum, granting "security" for the market. Second, the reputation system should make it so that every "good" agent will not be not misrecognized as a "bad" one, ideally having the reputation level computed for the good agent corresponding to actual quality of their service so the "equity" on the market is preserved as well.

| | good/bad volume ratio | Not using reputation system | | | | | | | | | If using reputation system | | | | | | | | | |
|---|---|---|---|---|---|---|---|---|---|---|---|---|---|---|---|---|---|---|---|---|
| | | profit from scam, % | loss to scam, % | full norm | log ratings | downrating | precision | default | conservatism | decayed | profit from scam, % | loss to scam, % | Pearson average | Pearson latest | avg good acc. | avg bad acc. | avg bal. acc. | avg good rmsd | avg bad rmsd | avg rmsd |
| r_20_0.1 | 8 | 9% | 1.1% | TRUE | TRUE | FALSE | 0.01 | 0.5 | 0.5 | 0.0 | 9% | 1.1% | -1.00 | -1.00 | 0.11 | 0.00 | 0.06 | 0.89 | 1.00 | 0.91 |
| r_20_0.1 | 8 | 9% | 1.1% | FALSE | TRUE | FALSE | 0.01 | 0.5 | 0.5 | 0.0 | 9% | 1.1% | -1.00 | -1.00 | 0.84 | 0.01 | 0.43 | 0.16 | 0.99 | 0.47 |
| r_20_0.1 | 8 | 9% | 1.1% | FALSE | FALSE | FALSE | 0.01 | 0.5 | 0.5 | 0.0 | 9% | 1.1% | 1.0 | 0.54 | 1.00 | 0.01 | 0.50 | 0.00 | 0.99 | 0.44 |
| r_20_0.1 | 8 | 9% | 1.1% | FALSE | FALSE | FALSE | 0.01 | 0.9 | 0.1 | 0.0 | 9% | 1.1% | 0.99 | 0.53 | 1.00 | 0.01 | 0.50 | 0.00 | 0.99 | 0.44 |
| r_20_0.1 | 8 | 9% | 1.1% | TRUE | FALSE | FALSE | 0.01 | 0.5 | 0.5 | 0.0 | 5% | 0.7% | 0.63 | 0.59 | 0.78 | 0.84 | 0.81 | 0.41 | 0.16 | 0.37 |
| r_20_0.1 | 8 | 9% | 1.1% | TRUE | FALSE | FALSE | 0.01 | 0.1 | 0.5 | 0.0 | <u>3%</u> | <u>0.4%</u> | <u>0.59</u> | <u>0.55</u> | <u>0.77</u> | <u>0.87</u> | <u>0.82</u> | <u>0.45</u> | <u>0.13</u> | <u>0.41</u> |
| r_20_0.1 | 8 | 9% | 1.1% | TRUE | FALSE | FALSE | 0.01 | 0.9 | 0.5 | 0.0 | 5% | 0.6% | 0.64 | 0.60 | 0.80 | 0.84 | 0.82 | 0.39 | 0.16 | 0.36 |
| r_20_0.1 | 8 | 9% | 1.1% | TRUE | FALSE | FALSE | 0.01 | 0.9 | 0.1 | 0.0 | 3% | 0.4% | 0.48 | 0.43 | 0.65 | 0.90 | 0.77 | 0.57 | 0.10 | 0.51 |
| r_20_0.1 | 8 | 9% | 1.1% | TRUE | FALSE | FALSE | 0.01 | 0.9 | 0.9 | 0.0 | 9% | 1.1% | 0.99 | 0.97 | 0.94 | 0.61 | 0.77 | 0.07 | 0.39 | 0.18 |
| r_20_0.1 | 8 | 9% | 1.1% | TRUE | FALSE | FALSE | 0.01 | 0.5 | 0.5 | 0.5 | 7% | 0.9% | 0.99 | 0.85 | 0.92 | 0.68 | 0.80 | 0.08 | 0.32 | 0.16 |
| r_20_0.1 | 8 | 9% | 1.1% | TRUE | FALSE | FALSE | 1.00 | 0.5 | 0.5 | 0.0 | 5% | 0.7% | 0.63 | 0.59 | 0.78 | 0.84 | 0.81 | 0.41 | 0.16 | 0.37 |
| r_20_0.1 | 8 | 9% | 1.1% | TRUE | FALSE | FALSE | 0.001 | 0.5 | 0.5 | 0.0 | 5% | 0.7% | 0.63 | 0.59 | 0.78 | 0.84 | 0.81 | 0.41 | 0.16 | 0.37 |
| r_20_0.1 | 8 | 9% | 1.1% | TRUE | FALSE | TRUE | 0.01 | 0.5 | 0.5 | 0.0 | 1% | 0.1% | 0.93 | 0.31 | 0.90 | 0.29 | 0.59 | 0.11 | 0.71 | 0.33 |
| r_20_0.1 | 8 | 9% | 1.1% | TRUE | FALSE | TRUE | 0.01 | 0.9 | 0.1 | 0.0 | <u>1%</u> | <u>0.1%</u> | 0.99 | 0.32 | <u>0.93</u> | <u>0.43</u> | 0.68 | <u>0.08</u> | <u>0.57</u> | 0.26 |
| r_20_0.1 | 8 | 9% | 1.1% | TRUE | TRUE | TRUE | 0.01 | 0.9 | 0.1 | 0.0 | 1% | 0.1% | -1.00 | -0.99 | 0.13 | 0.00 | 0.06 | 0.87 | 1.00 | 0.90 |
| r_20_0.1 | 8 | 9% | 1.1% | FALSE | FALSE | TRUE | 0.01 | 0.9 | 0.1 | 0.0 | 1% | 0.1% | -1.00 | -0.25 | 0.85 | 0.00 | 0.43 | 0.15 | 1.00 | 0.47 |
| r_20_0.1 | 8 | 9% | 1.1% | FALSE | TRUE | TRUE | 0.01 | 0.9 | 0.1 | 0.0 | 1% | 0.1% | -1.00 | -0.99 | 0.79 | 0.00 | 0.39 | 0.21 | 1.00 | 0.49 |
| r_100_0.1 | 40 | 44% | 1.1% | TRUE | TRUE | FALSE | 0.01 | 0.5 | 0.5 | 0.0 | 44% | 1.1% | -1.00 | -1.00 | 0.12 | 0.00 | 0.06 | 0.88 | 1.00 | 0.91 |
| r_100_0.1 | 40 | 44% | 1.1% | FALSE | TRUE | FALSE | 0.01 | 0.5 | 0.5 | 0.0 | 44% | 1.1% | -1.00 | -0.99 | 0.86 | 0.01 | 0.43 | 0.14 | 0.99 | 0.46 |
| r_100_0.1 | 40 | 44% | 1.1% | FALSE | FALSE | FALSE | 0.01 | 0.5 | 0.5 | 0.0 | 44% | 1.1% | 1.00 | 0.99 | 1.00 | 0.08 | 0.54 | 0.00 | 0.92 | 0.41 |
| r_100_0.1 | 40 | 44% | 1.1% | TRUE | FALSE | FALSE | 0.01 | 0.5 | 0.5 | 0.0 | 4% | 0.1% | 1.00 | 1.00 | 0.98 | 0.97 | 0.98 | 0.02 | 0.03 | 0.02 |
| r_100_0.1 | 40 | 44% | 1.1% | TRUE | FALSE | FALSE | 0.01 | 0.1 | 0.5 | 0.0 | 4% | 0.1% | 1.00 | 1.00 | 0.98 | 0.97 | 0.98 | 0.02 | 0.03 | 0.02 |
| r_100_0.1 | 40 | 44% | 1.1% | TRUE | FALSE | FALSE | 0.01 | 0.9 | 0.5 | 0.0 | 8% | 0.2% | 1.00 | 1.00 | 0.98 | 0.97 | 0.98 | 0.02 | 0.03 | 0.02 |
| r_100_0.1 | 40 | 44% | 1.1% | TRUE | FALSE | FALSE | 0.01 | 0.9 | 0.1 | 0.0 | 4% | 0.1% | 1.00 | 0.99 | 0.98 | 0.97 | 0.98 | 0.02 | 0.03 | 0.02 |
| r_100_0.1 | 40 | 44% | 1.1% | TRUE | FALSE | FALSE | 0.01 | 0.9 | 0.9 | 0.0 | 29% | 0.7% | 1.00 | 1.00 | 0.97 | 0.95 | 0.96 | 0.03 | 0.05 | 0.04 |

Figure 2. Using the reputation system with fixed market conditions and varying reputation system parameters. Rows 1-17 corresponds to the case when transactions by good agents are *20* times more valuable, in general, and rows 18-25 – to the case when this ratio is *100* times. In both cases, the population of bad agents was *20%* and each bad agent had *10* times more transactions than any of the good agents. Columns 3-4 show "profit from scam" and "loss to scam" ratios for the case when the reputation system is not used by agents while columns 12-23 hold these ratios for the case when the reputation system is used with different sets of parameters indicated in columns 5-11. Columns 14-21 show parameters for Pearson, accuracy, and root-mean-square deviation. The coloring scheme used is the same as on Fig.1. The optimal ("winning") set of parameters is underlined at row 6, with balanced optimum across metrics.

The simulations have been run for three different market scales - *10* agents during *10* days, *100* agents operating during *3* months and *1000* agents operating during *6* months. For these scales, average transaction cost ratios between "good" and "bad" agents were explored at *10, 20* and *100*. All combinations of ratios between consumers and suppliers as listed above were explored as well. Also, different reputation system parameters were tried for the market scale of *1000* agents operating during *6* months and transaction cost ratios *20* and *100*. For every run, the following metrics were computed were computed and collected with selected examples shown on Fig.1 and Fig.2.

- $V_g/V_b$ - good/bad volume ratio for market
- $C_{gbb}$ - profit from scam

- $C_{gbg}$ - loss to scam
- *PCC* - Pearson correlation between expected and computed "goodness", averaged across period and latest at the end of the period
- $A_g$ - accuracy for "goodness"
- $A_b$ - accuracy for "badness"
- $A_m$ - mean accuracy for both "goodness" and "badness"
- $D_g$ – root-mean-square deviation for "goodness"
- $D_b$ – root-mean-square deviation for "badness"
- $D_m$ – overall root-mean-square deviation

Based on the studies ad numbers on Fig.1 and Fig.3 and similar results for other market conditions, the following observations were being consistent across different market scales and reputation system parameters.

- **Using non-weighted explicit ratings did not work** under any of the conditions that we tested, and so it was never possible to get any of the metrics A far from *0.0* and any of the metrics D close from *1.0*, which implies that traditional reputation systems may not work under such conditions. This was also backed up with the fact that the $C_{gbb}$ value for either using a reputation system or not using it were the same, so that neither the loss of good agents to scam nor the profit of bad agents from scam depended on the use of such a reputation system. Whether this holds true for real marketpances needs further studies with more realistic simulations matching specific market parameters, including the scamming populations percentage, transaction ratios, volume and amount ratios and rates of newcomers and leavers.
- **Using either financially weighted explicit ratings or implicit financial ratings alone makes it possible** to decrease "loss to scam" and "profit from scam" both. Given "healthy" market conditions it is also possible to get all *PCC* and *A* metrics close to *1.0* and *D* close to *0.0* with $C_{gbg}$ and $C_{gbb}$ decreased up to *10* times, depending on other market and reputation system parameters.
- **Using explicit ratings weighted by financial values works the best**, granting generally higher *PCC* and *A* metrics and lower *D* metrics and being able to grant noticeable increase in $C_{gbg}$ and $C_{gbb}$ values in both "healthy" and "semi-healthy" market conditions.
- **For target market conditions** of *90%* consumers and *10%* suppliers, for the case of 1000 agents acting during 6 months, regardless of the explored market volume parameters, **use of the reputation system makes it possible** to keep "loss to scam" ($C_{gbg}$) value under 1% and "profit from scam" ($C_{gbb}$) under *5%* and, decreasing them up to ten times for "semi-healthy" and "healthy" markets.
- **Use of the reputation system by agents effectively decreases precision of it** because it saves "good" agents from experiencing "bad" agents and making "negative" ratings less effective than if the consumers were selecting suppliers blindly. That implies that there is no way to secure everyone in a truly distributed reputation system, because some "good" agents have to suffer from "bad" ones in order to save the rest with their ratings. However, it makes it possible to minimize the level of overall suffering for "good" agents and make "bad" agents much less profitable.
- **There is a tradeoff between minimizing the level of overall suffering (granting "security") and ability to recognize well-performing but less popular agents (granting "equity")** in "unhealthy" and "semi-healthy" market conditions so that a higher $A_g$ and a lower $D_g$ are connected with a lower $A_b$ and a higher $D_b$, or the other way around, depending on the reputation system parameters. This implies that better recognition of "good" agents or better "equity" costs unrecognizing some "bad" agents while perfect "security" leads to worse "equity" so that some of the good agents are misclassified as "bad" ones.
- **The parameters** *Weighting=True, FullNorm=True, LogRatings=False* have turned out to be necessary for the reputation system to work across all tried market conditions. The other parameters had no impact on results in "healthy" conditions. However, for "semi-healthy" conditions, lower conservatism *C* has provided noticeably better "security" sacrificing the "equity" while higher default has provided slightly better "equity" and "security" both. In turn, increase of decayed reputation $R_c$ has substantially increased "equity" but has also made "security" substantially worse. Finally, use of *Downrating=True* made it possible to decrease $C_{gbg}$ and $C_{gbb}$ ("loss to scam" and "profit from scam") to the factor of *10*, compared to variations of all of the other parameters, making them *10* times less than in the case when the reputation system was not used, however the cost of that came in the form of a low $A_b$ and higher $D_b$, apparently because "bad" agents evicted from selection list quickly were not getting enough low ratings and were thus given the chance to appear in shortlist above 40% over and over again.
- Given the above analysis and the numbers on Fig 3., **the best combination** of parameters has appeared to be *FullNorm=True, Weighting=True, LogRatings=False, $R_d$=0.9, C=0.1, $R_d$=0.0*.
- A special case is provided with *Downrating=True* setting ,underlined at row 14 on Fig.2, which needs more studies in the future research under more realistic market simulations.
- Highly negative Pearson correlation (above *0.95*) has been found between *A* and *D* families of **metrics** which has suggested that only one of them may be used for assessment of the reputation system performance in the further research. Similarly, high positive Pearson correlation has been found between *PCC* and $A_m$ values suggesting that only the latter measure or its counterpart $D_m$ may be used because the *A* and *D* families make it also possible to evaluate performance for "security" and "equity" independently.
- Highly positive Pearson correlation (above *0.9* in case or reliably working reputation system in use) has been found between the $V_g/V_b$ ratio and *PCC* values, implying that the ability of the reputation system to grant "security" and "equity" both is bound on the "health" and descends as the scam starts to prevail in the environment.
- Positive Pearson correlation (around *0.5* in case of a reliably working reputation system in use) has been found between *PCC* and $C_{gbb}$, which is connected with correlation to $V_g/V_b$ which means that lesser fraction and volume of scamming transactions on the market and/or better accuracy of the reputation system provide greater the profitability for those few who still perform scamming behavior and manage to fool the reputation system.

## 4 Conclusion

Based on the scope of performed work, we can conclude the following.

- For the studied market conditions, use of reputation system based on explicit ratings weighted by financial values of committed transactions provides both "security" and "equity" for the honest participants of the market, which suggests a need for further research in this direction with more realistic simulations.
- Use of a traditional reputation system with explicit ratings, under studied conditions, could be misleading in such market conditions providing neither "security" and "equity". In our study, the use of financial costs as implicit ratings is almost as good as use of explicit ratings weighted by financial values, but still provides less "security" and "equity". Further research is required to confirm these claims applied to real cases and to find the borderline conditions when this really start taking place.
- For extremely unhealthy market conditions it may be more safe to not use a reputation system that relies on everyone's personal experience as the reputation system may be cheated if the volume of scam-driving transactions is comparable to the volume of fair transactions.
- The use of a reputation system has side effects such as a) losing accuracy in identifying scamming participants because they are not being experienced by users . b) having honest agents miscounted as scamming agents or having actual performance of the honest agents under-evaluated. Potentially, both tradeoffs may to be avoided with careful system parameter tuning and implementation of extra policies for the process of making choices of suppliers on behalf of consumers. More research is required in this direction.

Our further work will be dedicated to the following.

- Making the simulations more realistic, exploring larger scales of markets above 1000 agents and 6 months, studying wider ranges of possible market conditions with different populations of scamming agents, different ratios of transactions committed by the agents and implementing different temporal patterns of the behaviors of the agents, primarily having different rates of newcomers and leavers, and introducing the notion of real transactional costs so the profit from scam may become tangible.
- Evaluating the possibility of improving "security" and "equity" with the use of rating aggregation and with the simulation of the staking-based [Kolonin *et al*., 2018] reputations as well as exploration of its impact on improvement of "equity", in respect to newcomers - especially.
- Explore the possibility of an implementation of dynamic reputation threshold management that has homeostatic control of the system enabling dynamic control on the threshold value separating "most likely scam" and "most likely honest" based on the rates of overall customer satisfaction in the system, such as percentage of non-zero ratings , or average rating value for the past period.